# Deconvolution optical-resolution photoacoustic microscope for high - resolution imaging of brain

Xianlin Song [a, *, #], Lingfang Song [b, #], Anqing Chen [a, #], Jianshuang Wei [c, d]

[a] School of Information Engineering, Nanchang University, Nanchang 330031, China;
[b] Nanchang Normal University, Nanchang 330031, China;
[c] Britton Chance Center for Biomedical Photonics, Wuhan National Laboratory for Optoelectronics-Huazhong University of Science and Technology, Wuhan 430074, China;
[d] Moe Key Laboratory of Biomedical Photonics of Ministry of Education, Department of Biomedical Engineering, Huazhong University of Science and Technology, Wuhan 430074, China;

[#] equally contributed to this work
* songxianlin@ncu.edu.cn

**ABSTRACT**

Photoacoustic imaging is becoming a very promising tool for the research of living organisms. It combines the high contrast of optical imaging and the high resolution of acoustic imaging to realize the imaging of absorption clusters in biological tissues. Since the scattering of ultrasound signals in biological tissues is 2-3 orders of magnitude weaker than the scattering of light in biological tissues, the endogenous absorption difference of tissues is directly used in the imaging process, so photoacoustic imaging has the advantages of deep imaging depth and non-destructive. As an important branch of photoacoustic imaging, photoacoustic microscopy can provide micron-level or even sub-micron-level imaging resolution, which is of great significance for biological research such as blood vessel detection. Since the lateral resolution of the photoacoustic microscopy imaging system depends on the focus of the laser, a higher resolution can be obtained by increasing the numerical aperture of the condenser objective. However, a large numerical aperture usually means a shorter working distance and makes the entire imaging system very sensitive to small optical defects. Therefore, the improvement of resolution through this method will be limited in practical applications. This paper implements a method of using iterative deconvolution to obtain a high-resolution photoacoustic image of the brain. The focal spot of the photoacoustic microscopy is measured to obtain the lateral PSF (point spread function) of the system. Making the measured PSF as the initial system PSF to perform Lucy- Richardson (LR) deconvolution. The image resolution of cerebral vasculature obtained by this method is higher. The full width at half maximum (FWHM) of width of the same cerebral capillaries before and after deconvolution are 7 μm and 3.6 μm, respectively, and the image definition is increased by about 1.9 times. Experiments show that this method can further improve the clarity of photoacoustic images of cerebral capillaries, which lays the foundation for further research on brain imaging.

## 1. INTRODUCTION

Observation and manipulation of cortical vascular network is an important way to study the structure and function of cerebral cortex microvasculature. In recent years, the development of modern optical imaging technology and the emergence of various marking technologies have made it possible to observe changes in cortical vessel morphology and hemodynamic parameters at the body level. Through real-time, in-vivo observation and manipulation of cortical microvasculature, it is expected to reveal the biological laws of cortical microvasculature growth, development and microcirculation homeostasis regulation, and explain the development and pathogenesis of related brain diseases. And is conducive to the development of effective treatments.

However, the mixed pool characteristics of the skull above the cortex have severely restricted the application of optical imaging technology in the imaging of blood vessels and blood flow through the cortical blood vessels in vivo. Researchers often use the method of removing the skull to achieve dynamic monitoring of the structure and function of cortical microvasculature. The open surgery will inevitably bring negative effects, which greatly limits the research on the structure and function of the cortical microvasculature at the body level.

Photoacoustic imaging is a promising technique that combines optical contrast with ultrasonic detection to map the distribution of the absorbing pigments in biological tissues [1-4]. It has been widely used in biological researches, such as structural imaging of vasculature [5], brain structural and functional imaging [6], and tumor detection [7]. Considering the lateral resolution of photoacoustic microscopy (PAM), it can be classified into two categories: optical-resolution (OR-) and acoustic-resolution (AR-) PAM [8, 9]. In AR-PAM, the spatial resolution is determined by the acoustic focus, since the laser light is weekly or even not focused on the sample. Conversely, in the OR-PAM, the laser light is tightly focused into the sample to achieve sharp excitation. Since the scattering of ultrasound signals in biological tissues is 2-3 orders of magnitude weaker than the scattering of light in biological tissues, the endogenous absorption difference of tissues is directly used in the imaging process, so photoacoustic imaging has the advantages of deep imaging depth and non-destructive. Since the lateral resolution of the photoacoustic microscopy imaging system depends on the focus of the laser, a higher resolution can be obtained by increasing the numerical aperture of the condenser objective. However, a large numerical aperture usually means a shorter working distance and makes the entire imaging system very sensitive to small optical defects. Therefore, the improvement of resolution through this method will be limited in practical applications.

This paper implements a method of using iterative deconvolution to obtain a high-resolution photoacoustic image of the brain. The focal spot of the photoacoustic microscopy is measured to obtain the lateral PSF (point spread function) of the system. Making the measured PSF as the initial system PSF to perform Lucy- Richardson (LR) deconvolution. The image resolution of brain obtained by this method is higher. The full width at half maximum (FWHM) of width of the same cerebral capillaries before and after deconvolution are 9.52 μm and 6.65 μm, respectively, and the image definition is increased by about 1.4 times. Experiments show that this method can further improve the clarity of photoacoustic images of cerebral capillaries, which lays the foundation for further research on brain imaging.

## 2. DECONVOLUTION OPTICAL-RESOLUTION PHOTOACOUSTIC MICROSCOPE

### 2.1 System setup

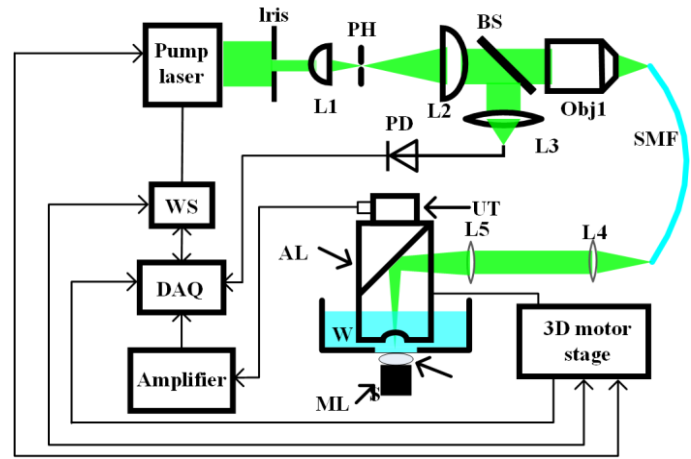

Figure 1. Scheme of the system. AL, acoustic lens; BS, beam sampler; DAQ, data acquisition card ; L1, L2, L3, L4 and L5, optical lenses; Obj1, objectives; PD, photodiode; PH, pinhole; S, sample; UT, ultrasonic transducer; W, water tank; WS, work station.

Fig. 1 shows the experimental setup of the optical-resolution photoacoustic microscope for in-vivo imaging of cerebral microvasculature. A dye laser (Credo, Sirah Laser und Plasmatechnik, Germany) pumped by a Nd:YLF laser (IS8II-E, EdgeWave GmbH, Germany) was employed as the wavelength-tunable irradiation source. It provided laser pulses with a pulse width of 9 ns and a pulse repetition of 1 KHz. Laser beam from the dye laser was focused by a set of plano-convex lens and then filtered by a 50-μm diameter pinhole. An objective lens was used to focus the laser beam into a single mode fiber (Core diameter: 3.5 μm). Laser beam from the distal end of the fiber was focused into the brain cortex by a long working distance objective lens (NT46-143, Edmund Optics, 4×, Working distance: 34 mm), after collimated by another same objective lens. The incident energy density on the brain cortex was estimated to be 18 mJ/cm$^2$, which was slightly smaller than the American National Standards Institute safety limit (20 mJ/cm2 in the visible spectral region).

A home-made prism which contained a silver gilt 45 °inclined plane was used for acoustic-optical coaxial alignment. The inclined plane reflected the incident laser beam, meanwhile a contact ultrasound transducer(V2022, Olympus) detected acoustic waves transmitted through the prism. An acoustic lens was ground in the bottom of the prism for acoustic focusing, and a plano-convex lens was inserted before the laser beam incident into the prism for correcting the divergence when laser beam passed through the acoustic lens. The prism we designed provided a more effective way to detect acoustic signal for it avoided the transformation from longitudinal acoustic wave to shear wave when the acoustic wave was reflected.

The imaging head was mounted on a three-axis linear stage, of which Z axis was used to adjust the focal plane. 3D imaging was implemented combining time-resolved acoustic detection and two-dimensional raster scan. The data acquisition card (ATS9350, Alazartech) recorded both photoacoustic signal and photodetector signal which indicated the deviation of incident laser energy, using a sample rate of 500 MHz. The lateral and axial resolution were estimated to be 4.07 μm and 14.4 μm, by imaging carbon nanoparticles with a mean diameter of 100 nm.

## 2.2 Blind deconvolution

For a linear time-invariant imaging system, the imaging process can be regarded as aconvolution of system point spread function (PSF) $h$ and the object $o$ mathematically. Hence, the inverse operation of convolution, i.e. deconvolution, can be used to restore the original object from the obtained blurred image $I$. To improve the contrast and resolution of the PAM MAP image, Matlab function "deconvblind" was applied to achieve blind deconvolution. The function employs maximum likelihood algorithm to estimate the optimal system PSF, through Lucy-Richardson (L-R) iteration processes. The focal spot of the photoacoustic microscopy (as show in Fig. 2) is measured to obtain the lateral PSF (point spread function) of the system. Making the measured PSF as the initial system PSF to perform Lucy- Richardson (LR) deconvolution.

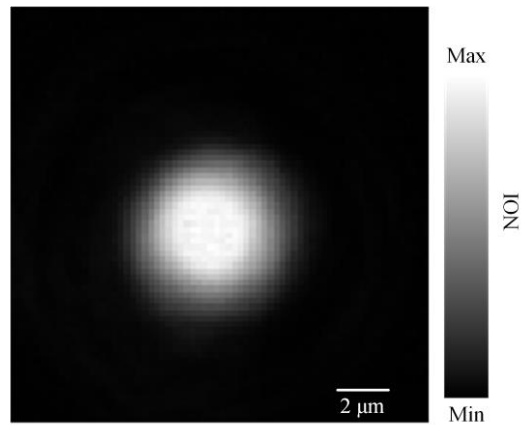

Figure 2. The focal spot of the photoacoustic microscopy. NOI, normalized optical intensity.

## 3. RESULTS

### 3.1 High -resolution imaging of brain

To fully demonstrate the performance of our multi-focus system, we performed in vivo cerebral vascular imaging on a mouse. A 20 g female Balb/c mouse was used to image. Before imaging, the mouse was anesthetized by intraperitoneal injection of chloral hydrate (0.2 g/kg) and urethane (1 g/kg), and then fixed on a brain stereotactic apparatus for further operation. Most of the skull was removed using a dental drill, forming a cranial opening window of 5 ×5 mm$^2$. After a layer of artificial cerebrospinal fluid had been implemented on the exposed dura mater, ultrasonic gel was applied for acoustic coupling.

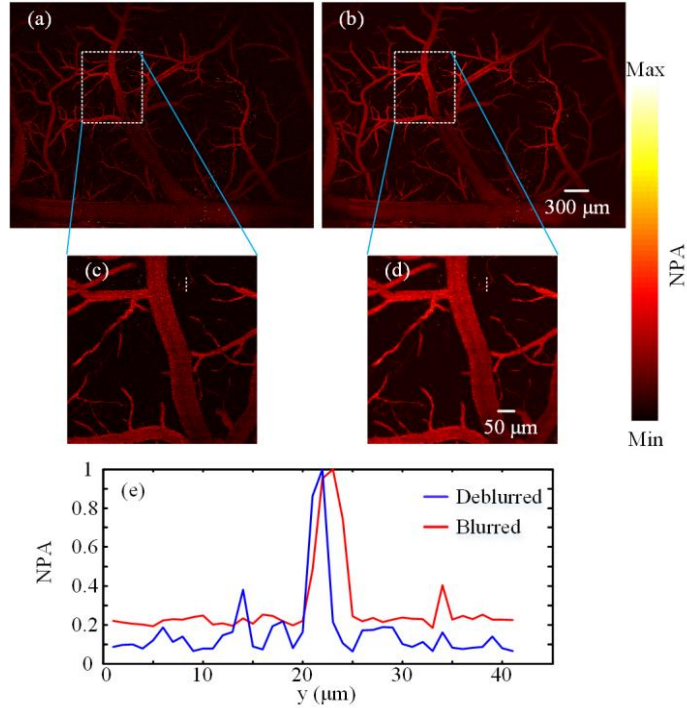

Figure 3. *In vivo* imaging of the mouse brain. (a) The deblurred MAP image by blind deconvolution of the mouse brain; (b) The blurred MAP image of the mouse brain. (c) and (d) are close-up images of the areas indicated by the white dashed rectangles in (a) and (b), respectively; (e) The normalized amplitude distribution along the dash line in (c) and (d). NPA, normalized photoacoustic amplitude.

As shown in Fig. 3, about 3 mm ×3 mm area of the open-skull mouse cerebral vasculature was imaged. Fig. 3(a) is the deblurred MAP image by blind deconvolution of the mouse brain. Fig. 3(b) is the blurred MAP image of the mouse brain. It can be seen that after using deconvolution, the cerebral blood vessels become clearer, more details can be shown. Fig. 3(c) and Fig. 3(d) are close-up images of the areas indicated by the white dashed rectangles in Fig. 3(a) and Fig. 3(b), respectively. Fig. 3(e) is the normalized amplitude distribution along the dash line in Fig. 3(c) and Fig. 3(d). The corresponding width of the vessel was defined as the FWHM of the distribution curve. It can be seen that before deconvolution, the width of blood vessel is 7 μm, and after deconvolution, the width of blood vessel is about 3.6 μm, and the resolution is improved by nearly 1.9 times.

## 4. CONCLUSION

We proposed a deconvolution optical-resolution photoacoustic microscope for high -resolution imaging of brain. The focal spot of the photoacoustic microscopy is measured to obtain the lateral PSF (point spread function) of the system. Making the measured PSF as the initial system PSF to perform Lucy- Richardson (LR) deconvolution. The image resolution of cerebral vasculature obtained by this method is higher. The full width at half maximum (FWHM) of the width of blood vessel before and after deconvolution are 7 μm and 3.6 μm, respectively, and the image definition is increased by about 1.9 times. Experiments show that this method can further improve the clarity of photoacoustic images of cerebral vasculature, which lays the foundation for further research on brain imaging.

# REFERENCES

[1] L. V. Wang, and S. Hu, "Photoacoustic Tomography: In Vivo Imaging from Organelles to Organs," Science 335(6075), 1458-1462 (2012).
[2] P. Beard, "Biomedical photoacoustic imaging," Interface Focus 1, 602-631 (2011).
[3] J. Yao and L. V. Wang, "Photoacoustic tomography: fundamentals, advances and prospects," Contrast Media Mol. Imaging 6(5), 332–345 (2011).
[4] L. V. Wang, "Prospects of photoacoustic tomography," Med. Phys. 35(12), 5758–5767 (2008).
[5] S. Hu and L. V. Wang, "Photoacoustic imaging and characterization of the microvasculature," J. Biomed. Opt. 15(1), 011101 (2010).
[6] X. Wang, Y. Pang, G. Ku, X. Xie, G. Stoica, and L. V. Wang, "Noninvasive laser-induced photoacoustic tomography for structural and functional in vivo imaging of the brain," Nat. Biotechnol. 21(7), 803–806 (2003).
[7] H. F. Zhang, K. Maslov, G. Stoica, and L. V. Wang, "Functional photoacoustic microscopy for high-resolution and noninvasive in vivo imaging," Nat. Biotechnol. 24(7), 848–851 (2006).
[8] Y. Liu, C. Zhang, and L. V. Wang, "Effects of light scattering on optical-resolution photoacoustic microscopy," J. Biomed. Opt. 17(12), 126014 (2012).